\magnification=\magstep1
\font\bigbfont=cmbx10 scaled\magstep1
\font\bigifont=cmti10 scaled\magstep1
\font\bigrfont=cmr10 scaled\magstep1
\vsize = 23.5 truecm
\hsize = 15.5 truecm
\hoffset = .2truein
\baselineskip = 14 truept
\overfullrule = 0pt
\parskip = 3 truept
\def\frac#1#2{{#1\over#2}}


\topinsert
%
\vskip 1.7 truecm
\endinsert
%
\centerline{\bigbfont MESOSCOPIC TRANSPORT AS MANY-BODY PHYSICS$^*$}
\footnote{}{* To be published in {\it Proceedings of the 25th
International Workshop on Condensed Matter Theories (CMT25),
Canberra 2001.}
}
\vskip 20 truept

\centerline{\bigifont Frederick Green}
\vskip 8 truept
\centerline{\bigrfont Centre for Quantum Computer Technology}
\vskip 2 truept
\centerline{\bigrfont School of Physics,
The University of New South Wales}
\vskip 2 truept
\centerline{\bigrfont Kensington NSW 2052, Australia}
\vskip 2 truept
\centerline{\bigrfont and}
\vskip 2 truept
\centerline{\bigrfont Department of Theoretical Physics}
\vskip 2 truept
\centerline{\bigrfont Research School of Physical Sciences and Engineering}
\vskip 2 truept
\centerline{\bigrfont The Australian National University}
\vskip 2 truept
\centerline{\bigrfont Canberra ACT 0200, Australia}

\vskip 8 truept
\centerline{\bigifont Mukunda P. Das}
\vskip 8 truept
\centerline{\bigrfont Department of Theoretical Physics}
\vskip 2 truept
\centerline{\bigrfont Research School of Physical Sciences and Engineering}
\vskip 2 truept
\centerline{\bigrfont The Australian National University}
\vskip 2 truept
\centerline{\bigrfont Canberra ACT 0200, Australia}

\vskip 14 truept

\vskip 1.8 truecm

\centerline{\bf 1. INTRODUCTION}
\vskip 12 truept

Mesoscopic transport physics has grown astoundingly in the last decade.
This kind of physics continues to astonish by the delicacy of its
architectures, and of the experiments done on them. In every way,
mesoscopics provides an ideal test-bed for many-body theory, since
the reduced dimensionalities and small scales naturally
amplify the effects of inter-electron correlations.
However, there is a deeper reason. It rests with the nature of
electronic transport and its microscopic origin in electron-hole
fluctuations.

The sustained development of this field
depends, therefore, on a matching capacity to understand dynamic
electron processes at length scales reaching down to the atomic.
Aside from its intrinsic scientific value,
such an understanding is crucial to the commercial
design and integration of novel electronic devices.

A standard bulk-averaging analysis will break down when actual
device sizes approach the typical mean free paths for scattering,
or the natural screening length.
In such systems, much more care has to be taken.
In the last two decades a body of work, widely seen as
having its origins in Landauer's early insights
[1],
has enjoyed great success in mesoscopic transport. The 
best known result within this approach is the Landauer formula
for the quantized conductance of a one-dimensional wire.

To many practitioners of mesoscopics, the Landauer formula seems
to possess quasi-magical powers in that it claims to explain  
a staggering diversity of transport measurements (as witness
the authoritative list in Reference [2]).
It is a formula that, in its self-confessed simplicity, bases
itself strictly on {\it one-body} physics and nothing else. As
a result, it contains none of the important electron
correlations that come into prominence for low-dimensional
mesoscopic structures. More serious is the fact that Landauer's
model ignores a series of indispensable physical bounds on the
behavior of the electron gas (see Sections 2 and 3 below for details).

A further important point is that many would-be proofs of Landauer's
popular formula insist upon the paradoxical notion that two
(or even more) chemical potentials, each thermodynamically distinct,
should coexist within a single, closed, electrical circuit when it
is out of electro-chemical equilibrium. On that account also,
Landauer's prescription falls well short of internal consistency.
We will deal with this issue on the way to setting out
a proper, microscopic view of small-scale transport.

For a good general reference on mesoscopics, see
the book by Ferry and Goodnick [1].
On the status of mesoscopic noise theory as popularly received,
refer to Blanter and B\"uttiker [2]. We also mention the
recent critiques of Landauer-B\"uttiker-Imry theory in the
review of mesoscopic transport by Agra\"{\i}t {\it et al.} [3].

In the evolution of this new branch of condensed-matter physics,
concepts of small-scale current behavior cannot be divorced from
their roots within microscopic electron correlations. These roots
are firmly grounded in the theory of charged Fermi liquids
[4--6].
Within that long-established canon, we have two aims:

\item{$\bullet$}
to help motivate mesoscopics as an inherently
{\it many-body} discipline, and

\item{$\bullet$}
to encourage a reciprocal interest in the fresh opportunities
for testing many-body ideas, as offered by mesoscopic physics.

\noindent
The relationship between microscopic
many-body physics and mesoscopic device physics is indeed two-way.
Not only does the former hold the foundations for a true understanding
of the latter; mesoscopics, in turn, manifests (often quite graphically)
the action of the conservation laws.
It is the miscroscopic conservation laws that give many-body theory
its universal reach and its great explanatory power
[7].

Attention falls on a primary conservation law: gauge invariance.
On it, one can build a firm and trustworthy picture of
dynamic fluctuations in an already classic mesoscopic system:
the one-dimensional metallic wire, or quantum point contact (QPC).
This case study highlights the
need for {\it any} mesoscopic transport model to be,
first and genuinely, a many-electron theory. It also points to the
prospect of researching strong correlations through
nonequilibrium current noise, a tool
that so far has been unfamiliar to many-body physics.

In Section 2 we review the general conditions applying to
an open and externally driven system of mobile electrons,
metallic ones in particular. Our discussion recalls a
series of insightful, if less often discussed, microscopic
analyses. These have been freely available
in the literature over the last decade
[8--13].
After that, in Sec. 3 we analyze the compressibility of
an open conductor. This parameter, closely allied to the
electron-hole fluctuations, is an exact invariant of the
system. It is absolutely unaltered by nonequilibrium transport.
This result extends beyond the free electron gas, to
systems that are subject to strongly inhomogeneous Coulomb screening.
Nonuniform electronic conductors are encountered everywhere
in mesoscopic physics, and internal screening leads to
the substantial suppression of their compressibility.
This effect too is fully inherited by a system
when it is driven out of equilibrium.
Last, we foreshadow the effect of exchange-correlation
renormalization on the compressibility. 

Charge conservation governs the compressibility
of a charged Fermi liquid. It also governs its response
right through the high-field nonlinear regime.
Section 4 retraces our kinetic noise theory with reference
to the critical QPC noise experiment of Reznikov {\it et al.}
[14],
whose features raise some outstanding issues of principle.
These are resolved in a natural way
within our kinetic many-body perspective, and compressibility
is the key to their resolution. We conclude in Sec. 5.

\vskip 28 truept 
\centerline{\bf 2. TRANSPORT IN OPEN ENVIRONMENTS}
\vskip 12 truept

A fundamental problem in mesoscopic transport is the
explicit consideration of what it means, in microscopic terms,
to address ``open environments'' that no longer seem to admit
direct treatment within a closed Hamiltonian formalism. 
Over the last decade, a number of important comprehensive papers
has appeared. Taken as a whole, they contribute substantially
towards the realization of a logical and calculable microscopic
approach to transport at mesoscopic scales.
This is so not just in linear response, but also in truly
nonequilibrium situations. 
The authors responsible for this body of work are:

\item{} Sols [8];
\item{} Fenton [9,10];
\item{} Magnus and Schoenmaker [11,12]; and
\item{} Kamenev and Kohn [13].

\noindent
We first examine the fundamental principles that they
articulate, dissect, and in large measure resolve.
It hardly needs saying that their critiques of mesoscopic
theory stand upon a time-tested and rigorous
analytic tradition, from Maxwell and Boltzmann through Fermi and Landau.

Neither is it a surprise to find that such first-principles
descriptions of the mesoscopic electron gas
begin -- and end -- with charge conservation.
However, its practical outworking can be subtle.
For instance, unlike an ideally closed system of
carriers not subject to exchange with the outside,
the global charge neutrality of a
real -- open -- conductor is not automatic,
so that a nontrivial demonstration is required.
Nonetheless, the physics of charge and number
conservation applies as strictly as ever.

\vskip 8 truept
\centerline{\it a. Gauge Invariance}
\vskip 8 truept

{\it Sols} [8] first established that an open metallic
conductor is globally gauge invariant if, and only if, the
microscopic equation of motion for the electrons
{\it explicitly} includes the contribution from each
source and sink of current in the problem. These are located at the
device's interfaces with the outlying reservoirs.
The reservoirs themselves remain in (local) equilibrium.

Sols' formalism introduces a critical separation of physical
roles within the mesoscopic description. Thus,
the dynamical action of charge injection and removal (which
brings about the system response) is entirely separate
from the equilibrating action of the reservoir leads.
The macroscopic leads act, in the first instance, not
as regulators of the current but to pin and stabilize
the nonequilibrium state of the intervening device.
{\it The current is not conditioned by the lead equilibria.}

The clear implication of Sols' examination is that
the current sources/sinks operate dynamically
in their own right. They have nothing to do with the
locally unchanging state of the reservoirs. According to
Ref. [8], injection and extraction must act,
and be represented, {\it manifestly}.

Transport descriptions that take no notice
of the entry and exit of flux are necessarily faulty.
Ref. [8] shows that the sources and sinks are
irreducible elements in a gauge-invariant model.
Thus it is theoretically not possible to cut them out of
open-system transport merely at convenience,
as if these essential flux exchanges could
be equated to a remote (and forgettable) background effect
[15].

From Sols, a major point follows:

\item{$\bullet$} {\it A microscopically conserving model
of transport and fluctuations must manifestly account for
all influx and efflux of current at the open boundaries.}

\noindent
Let us now focus on the quiescent reservoirs. They lie
outside the conducting sample, in which the interesting
transport phenomena all occur. At the same time, they are
in intimate electrical contact with it. As we have remarked,
Ref. [8] shows that the reservoirs' microscopic function is
distinct from the supply and recovery of the
(nonequilibrium) current fluxes.

Therefore the reservoir leads have two dominant properties,
which are {\it not} dynamic but, rather, thermodynamic.
The macroscopic leads are

\item{(i)} always neutral, and

\item{(ii)} always in local equilibrium.

\noindent
In other words, they pin the carriers asymptotically (via the
local Fermi levels in the respective leads,
as well as the thermodynamic bath temperature)
to their permanent and locally unchanging equilibrium state.
The corollary is that

\item{$\bullet$} {\it The nonequilibrium carrier states must
connect seamlessly to each invariant equilibrium state that
locally and uniquely characterizes each of the leads}.

\vskip 8 truept
\centerline{\it b. Screening}
\vskip 8 truept

Shortly after Sols' formal proof, {\it Fenton} [9,10] made a
detailed quantum-kinetic analysis of mesoscopic transport.
He highlighted the dominance
of charge neutrality of the reservoirs.
Through Thomas-Fermi screening, the asymptotic carriers
{\it confine} the electromotive force (EMF) so that it
is spatially coextensive with the driven conductor, across
whose boundaries it appears.

Fenton drew a deep conclusion from his first-principles
quantum analysis.
The quantized conductance observed in one-dimensional (1D) wires
is explainable if {\it and only if} its underlying model includes the
screening response of the reservoirs at the sample interfaces.
This is, of course, a consequence of Gauss' theorem applied
to the electron gas (the perfect-screening sum rule [4],
also recalled by Sols [8]).
It entails the statement that

\item{$\bullet$} {\it The interface regions of an open
mesoscopic sample -- the actual sites of the carriers'
physical transition to the asymptotic charge-neutral state --
must be included as part of the
theoretical description of the transport.}

\noindent
As in every electron-gas problem, screening is an integral
part of the procedure here.
Unless it operates, Landauer's quantized mesoscopic conductance
can find no reconnection with microscopic quantum kinetics
[9].
Fenton's analysis requires that -- rather than the common
practice of demoting metallic-electron screening to some
{\it ad hoc} perturbative afterthought
[15] -- a mesoscopic transport description make
such collective screening its centerpiece.

\vskip 8 truept
\centerline{\it c. Hamiltonian Description}
\vskip 8 truept

So far we have restated the prime roles of gauge invariance,
global neutrality, and equilibrium in charged open systems. This is
sufficient to set up a kinetic theory in the mesoscopic
regime, not only in the weak-field limit but well away from equilibrium
[16,17].

Since energy dissipation is a real and central effect in such cases,
it is easy to conclude that a canonical Hamiltonian approach is not
feasible. To the extent that a (dissipative) dynamical equation
cannot be generated from a standard hermitian Hamiltonian form,
one might think that the kinetic description of real carrier
dynamics, with its energy losses,
is destined to remain formally incomplete.

This incompleteness is not inevitable. A central fact
that should be taken into account is the
multiply connected topology of a closed circuit. 
An elegant microscopic formalism
for nonequilibrium, energy-dissipating and
closed mesoscopic circuits has been proposed by
{\it Magnus and Schoenmaker}
[11,12].
They point out that the toroidal topology of the
circuit (this includes the battery) admits valid
solutions to Maxwell's equations which, though locally
conservative, are {\it globally nonconservative}
along complete paths around the hole of the torus.

Magnus and Schoenmaker construct the gauge-invariant
Hamiltonian for a closed and {\it multiply connected}
system of driven carriers. The classic minimal-coupling
expression for the electronic Hamiltonian
is faithfully maintained. What is novel is that the
physical solution, resulting from the geometrical
reconnection, will dissipate energy from the battery
along any full circuit path.

In this way, real processes can be described
without altering the locally conservative Hamiltonian structure
of the interacting electrons. All of the powerful quantum kinetics of
Kadanoff-Baym [18], or of Keldysh [19],
becomes immediately applicable in principle.

Under very general conditions, the authors formally
prove that the external power $P$ supplied by the source is
dissipated according to the standard formula $P = IV$,
in terms of circuit current $I$ and EMF $V$. The result
is intuitively clear, but challenging to prove. The
derivation is fully quantal, and valid at any device scale.

The work of Magnus and Schoenmaker dovetails with that of Sols
and Fenton. The treatments are all equivalent in the end;
the latter two stress the governing role of
the boundary conditions within a
charge-conserving description of a finite mesoscopic device
that is {\it electrically open to the outer leads}.
On the other hand, the former approach restates the critical
fact that any circuit path containing the device
must be {\it closed}, and that the topology of the closure
makes possible (i) long-range neutrality, (ii) dissipation, and hence
(iii) dynamical stability within a unified Hamiltonian description.

\vskip 8 truept
\centerline{\it d. Uniqueness of the Chemical Potential}
\vskip 8 truept

Finally, the recent many-body derivation of Landauer's conductance
formula by {\it Kamenev and Kohn} [13] is noteworthy.

\item{(a)} It sets
the Landauer result within a solidly orthodox quantum-kinetic
formulation, free of the unnecessary phenomenology that has weakened
previous such proofs;

\item{(b)} it provides a detailed and self-contained treatment
of collective screening;

\item{(c)} by applying Kubo response theory in its appropriate
form, Ref. [13] naturally preserves the logical structure of
the microscopic fluctuation-dissipation theorem;

\item{(d)} it demonstrates the unphysical character of a purely
heuristic scheme now widely accepted in mesoscopics:
namely, the practice of ``driving'' all
currents by a kind of pseudodiffusive process, generated by
a virtual mismatch in the chemical potential across
various leads.

\noindent
At the two-body level, a consequence of Refs. [8-12]
is to rule out that last artifice in all its variations,
for they all lead inevitably to the violation of gauge invariance
[20].

It is Kamenev and Kohn's work
[13],
however, that drives home the fictional nature of
pseudodiffusive phenomenology, by straightforward
{\it ab initio} evaluation of Kubo's conductance
formula for a 1D wire. Just one parameter is needed:
the equilibrium chemical potential. Thus

\item{$\bullet$} {\it the microscopic properties of a
mesoscopic system are uniquely determined by
one chemical potential, and one only:
that of the reference equilibrium state.}

\vskip 8 truept
\centerline{\it e. Microscopics and Phenomenology}
\vskip 8 truept

Each of the approaches enumerated above counters
a viewpoint commonly advanced by other models
[2,15,21,22].
As already
mentioned, the electrical currents in such phenomenologies
are generated and (allegedly) sustained by passive connection of the
mesoscopic device to two or more large reservoirs
at unequal potentials. Instead
of being allowed their correct thermodynamic values,
measured from the bottom of their occupied {\it local} conduction bands,
the chemical potentials in the reservoirs are taken
from a global zero of energy
which itself has no unique physical significance.
This immediately breaks the canonical
linkage that ties together the {\it physical}
carrier density and the {\it physical} compressibility
of the collective system
[20].
We return to the compressibility below.

In the dominant conception of mesoscopics,
carriers are never driven dynamically through
the conductor, from one lead to another, by the real field.
Rather, they fall passively, and by sheer happenstance,
over the associated Fermi-level drop.
All that can be known of the dynamics are the
asymptotic equilibrium states, and
the asymptotic probability
for a single carrier to drop over the potential cliff.

This picture is of no help in explaining the
{\it energy dissipation} that has to accompany
open-system transport.
Pseudodiffusive phenomenologies have no access to
anything but generic equilibrium properties.
From a detailed kinetic perspective, those models
cannot contribute to understanding the critical
nonequilibrium processes unfolding within the conductor,
and at its interfaces with the reservoirs.
These processes dissipate energy. Dissipation
is out of scope.

\topinsert
\input psfig.sty
\vskip 0.25truecm
\centerline{\hskip10mm\psfig{figure=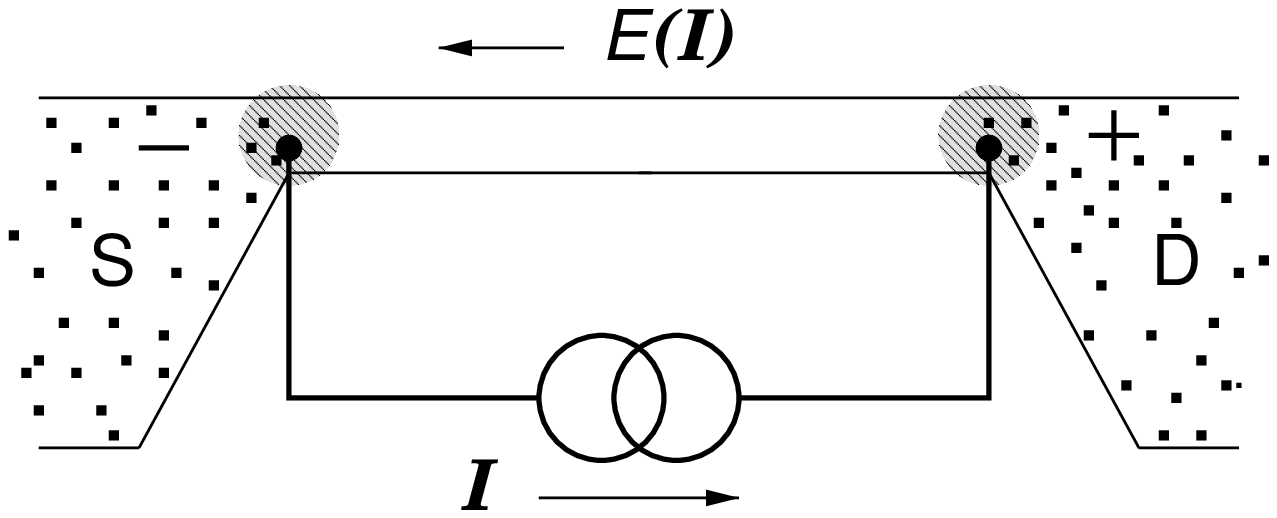,height=3.5truecm}}
\vskip 0.55truecm 

\noindent
{\bf Figure 1.}
Mesoscopic transport in a nutshell.
A closed, i.e. toroidal, loop includes the conductor under test
(clear central area).
The loop is the essential supplier of the current $I$, either
by a generator as shown or, equivalently, by the EMF from a battery.
The source (S) and drain (D) reservoirs ensure
(a) {\it inelastic dissipation} of the electrical power dumped to
the circuit, (b) {\it screening} of all fields including
the driving field $E(I)$ across the nonequilibrium region,
and (c) {\it continuity} of the carrier distribution.
These dynamically competing processes are located at
the interfaces (shaded areas).
The boundary conditions illustrated here are necessary
and sufficient for a microscopically conserving description.
(This Figure is taken from our Ref. [23]).
\vskip 8truept
\endinsert
\hsize = 15.5 truecm

Figure 1 summarizes the physical structure of a mesoscopic
system open to its macroscopic electrical surroundings,
within which all measurements take place. The closed loop
supplies the flux, as specified in Ref. [11] while,
in accordance with Ref. [8], the flux entry and exit points
maintain global conservation as perceived by a test charge
{\it within the wire}. The large reservoirs maintain
thermodynamic stability by pinning the local Fermi level
at the interfaces, by providing energy dissipation (Ref. [11]),
and by confining the electromotive forces via screening
(Ref. [9]).

By simple logic, the phenomenological scenarios cannot
allow the reservoirs
themselves to reconnect electrically ``at infinity'' and
thus to close the circuit. If they did so, the
resulting coexistence of several {\it different} chemical potentials
(required for pseudodiffusive ``transport'')
would pose an immediate consistency problem [11,13].
There is no flux generator, and thus
no replenishment of the ever-growing charge
imbalance between the reservoirs.
Therefore an adiabatically sustainable, current-carrying
steady state cannot be described within such an approach.

By contrast, in a standard many-body description
all the dynamic and dissipative effects that determine
both transport and noise conform to the microscopic
conditions articulated in Refs.
[8--13],
and summarized in Fig. 1.
Without those normative boundary conditions
(which include the batteries and/or current generators),
a well-controlled model cannot be set up.
That is the message of the works we have recalled.

\vskip 28 truept 
\centerline{\bf 3. A CANONICAL SUM RULE: COMPRESSIBILITY}
\vskip 12 truept

As is well known, the physics of the electron gas is
strongly conditioned by a set of canonical theorems,
the sum rules
[4].
Several of the sum rules are directly linked to number
conservation, and thus to gauge invariance. They,
if nothing else, should be identically satisfied by any
model of transport and fluctuations.

The electron-gas sum rules share a common template.
On the left-hand side, a physical quantity is given as
a function of the thermodynamic and {\it single-particle}
parameters (temperature; chemical potential; density etc.).
This is equated, on the right-hand side,
to a {\it two-particle correlation function}
[24].

Conceptually, the sum rules are closely related to the
fluctuation-dissipation theorem
[25,26];
once again we see a one-body parameter,
the electronic conductance, on one side of the relation.
On the other side, it is uniquely determined
by an irreducible two-body
response function: the current-current correlator.
Equivalences such as these express the microscopic conservation
of energy, momentum, and particle number at each elementary
electron-hole vertex in the system's quantum many-body expansion
[5].

With this background, let us examine the content and
mesoscopic consequences of the simplest of the sum rules, that for
compressibility. On its own, that rule is a surprisingly
powerful constraint on the behavior of mesoscopic conductors.
For the full and formal details, see our Refs. [16,17,20].

\vskip 8 truept
\centerline{\it a. The Open Electron Gas}
\vskip 8 truept

Consider a driven, metallic conductor of volume $\Omega$
bounded by {\it equilibrated} and {\it charge-neutral}
reservoirs. Here we invoke the boundary conditions discussed
above; see Sec. 2.a. That the well-defined active region,
including the interfaces, must be fixed and finite follows from
Fenton's microscopic analysis; see Sec. 2.b.

Within the device, let $N$ be the number of mobile carriers,
and let $\Delta N \sim {\langle N^2 - {\langle N \rangle}^2 \rangle}$
be the mean-square fluctuation of $N$, where the averaging will
be made precise shortly. Since global neutrality holds,

\item{$\bullet$} $N$ {\it is unconditionally invariant for
all transport of charge through the system}.

\noindent
The total number $N$ can be accounted for
in two ways. We can integrate the
time-dependent nonequilibrium distribution $f_{\alpha}(t)$
for the configuration-space label $\alpha = ({\bf r}, {\bf k})$
(spin and many-valley indices are implied). Alternatively,
we can add up all the electrons in $\Omega$
for the equilibrium case, where the one-body
distribution is

$$
f^{\rm eq}_{\alpha} =
{1\over {1 + \exp{[(\varepsilon_k + U({\bf r}) - \mu)/k_{\rm B}T]} }}.
$$

\noindent
The (quasiparticle) kinetic energy is $\varepsilon_k$; the
Hartree mean-field potential is $U({\bf r})$, and is generally a
functional of the density distribution. The global
chemical potential $\mu$ is  fixed by the outer reservoirs.

Very simply,

$$
\sum_{\alpha} f_{\alpha}(t) = N = \sum_{\alpha} f^{\rm eq}_{\alpha}.
\eqno(1)
$$

\noindent
The origin and meaning of this equation are clear. We
add up all contributions to $N$
over the space of wavevectors ${\bf k}$
and over all ${\bf r}$ within the real-space
volume $\Omega$. The operation $\sum_{\alpha}$
denotes this configurational integration.
Since the region of physical interest is fixed and
unconditionally neutral, $N$ is always exactly balanced
by the unchanging positive background. Equation (1) follows.

We now sketch the corresponding fluctuation relation.
It is a consequence of Gauss' law, expressed through Eq. (1).
The {\it difference function}

$$
g_{\alpha}(t) = f_{\alpha}(t) - f^{\rm eq}_{\alpha}
$$

\noindent
measures the local out-of-balance component of the ensemble.
Eq. (1) is equivalent to

$$
\sum_{\alpha} g_{\alpha}(t) = 0.
\eqno(2)
$$

\noindent
If we take variations of the difference function,
for example with respect to the
chemical potential proper to the reservoir leads (this
is done by changing the leads'
asymptotic electron density {\it and that of
its neutralizing positive background}), then

$$
\sum_{\alpha}
{\partial g_{\alpha}(t)\over \partial \mu} \equiv
\sum_{\alpha}
{\left(
\sum_{\alpha'} {\delta g_{\alpha}(t)\over \delta f^{\rm eq}_{\alpha'}}
{\partial f^{\rm eq}_{\alpha'}\over \partial \mu}
\right)}
= 0.
\eqno(3)
$$

\noindent
We have used the variational chain rule to partly unpack the
rich inner
structure of this fluctuation; the functional derivative
$\delta g_{\alpha}(t)/\delta f^{\rm eq}_{\alpha'}$ is a specialized
Green function whose form is calculable from the
kinetic equation for $f(t)$
[17].
It is the centerpiece of our approach.
\vskip 8 truept
\centerline{\it b. Free-Electron Compressibility}
\vskip 8 truept

Equation (3) is the cardinal element in deriving
the nonequilibrium compressibility sum rule.
First, the equilibrium mean-square fluctuation
in local carrier number is given by

$$
\Delta f^{\rm eq}_{\alpha}
\equiv k_{\rm B}T {\partial f^{\rm eq}_{\alpha}\over \partial \mu}
= f^{\rm eq}_{\alpha}(1 - f^{\rm eq}_{\alpha}),
\eqno(4a)
$$

\noindent
which lets us compute the total mean-square fluctuation in $N$:

$$
\Delta N \equiv k_{\rm B}T {\partial N\over \partial \mu}
= \sum_{\alpha} \Delta f^{\rm eq}_{\alpha}.
\eqno(4b)
$$

\noindent
Second, the exact {\it nonequilibrium} mean-square fluctuation is
[17]

$$
\Delta f_{\alpha}(t) = \Delta f^{\rm eq}_{\alpha}
+ \sum_{\alpha'} {\delta g_{\alpha}(t)\over \delta f^{\rm eq}_{\alpha'}}
\Delta f^{\rm eq}_{\alpha'}.
\eqno(5)
$$

\noindent
From Eq. (3) we get, at last,

$$
\sum_{\alpha} \Delta f_{\alpha}(t) = \Delta N
= \sum_{\alpha} \Delta f^{\rm eq}_{\alpha}.
\eqno(6)
$$

The compressibility for an
undisturbed free-electron gas is standard
[4]:

$$
\kappa \equiv {\Omega\over N^2} {\partial N\over \partial \mu}
= {\Omega\over N k_{\rm B}T} {\Delta N\over N}. 
\eqno(7)
$$

\noindent
Note that, so far, we have taken our variations (and the
associated fluctuation structures) with all internal fields
held {\it fixed}; that is, we freeze the potential $U({\bf r})$.
Shortly, we will relax this constraint. At this stage our
treatment is precisely analogous to the Lindhard model of
the electron gas.

Our result is immediate. Eqs. (1) and (6) are exact within
the nonequilibrium kinetic description. Therefore

\item{$\bullet$}{\it The invariant compressibility of
an open mesoscopic
conductor, driven at any current, is given by Eq. (7).}

\noindent
A simple feature of Eq. (7) is seen in its Maxwellian
limit. In that case, $\Delta f^{\rm eq} = f^{\rm eq} \propto
e^{\mu/k_{\rm B}T}$. The ratio $\Delta N/N$ is unity.
That renders the right-hand expression equal to the
inverse of the pressure
$P = Nk_{\rm B}T/\Omega$. Eq. (7) then states the
classical, ideal-gas result

$$
\kappa_{\rm cl} = P^{-1}.
$$

\noindent
When the system is degenerate, then $\Delta f^{\rm eq} < f^{\rm eq}$
and

$$
\kappa_{\rm free} = \kappa_{\rm cl} {\Delta N\over N}
< \kappa_{\rm cl}
\eqno(8)
$$

\noindent
(we denote the free ``Lindhard'' compressibility by
$\kappa_{\rm free}$ in anticipation of the next subsection). 
This shows the stiffening action of degeneracy on the system,
in loose analogy with van der Waals' concept of the
excluded hard-core volume. The effect clearly persists,
{\it with no change at all}, out of equilibrium. This is
despite the fact that the underlying microscopic
form of the dynamic $\Delta f(t)$ is far more complex than
its generic equilibrium limit, $\Delta f^{\rm eq}$.
The power of the compressibility sum rule comes precisely
from its total, and surprising, insensitivity to external
driving forces. 

We have just exhibited a general result for nonequilibrium
electronic conductors. It is the direct consequence of conservation,
acting in the context of the boundary conditions of Sec. 2.
These boundary conditions are the only ones fully consistent
with microscopic principles that hold, universally, in a
many-body situation. The mesoscopic electron gas is one such.

\eject
\vskip 8 truept
\centerline{\it c. Screened Compressibility}
\vskip 8 truept

It is time to focus on the reaction of the internal fields
to the carrier fluctuations. Intuitively, any such
fluctuation costs additional electrostatic energy.
The system will minimize the excess internal energy
by self-consistent screening of the initial charge
fluctuation. This is Coulomb-induced suppression,
of which the Coulomb blockade
is one of the more dramatic instances.

Let us start with the self-consistent equilibrium
response:

$$
k_{\rm B}T{\delta f^{\rm eq}_{\alpha}\over \delta \mu}
= {\delta\over \delta \mu}{\Bigl(\mu - U({\bf r}) \Bigr)}
k_{\rm B}T{\partial f^{\rm eq}\over \partial \mu}
=
{\left( 1 - {dU\over dn}{\delta n({\bf r})\over \delta \mu} \right)}
\Delta f^{\rm eq}_{\alpha}.
\eqno(9a)
$$

\noindent
We have made the usual Ansatz for $U({\bf r})$;
one assumes that it depends directly on the local
density so that $U \equiv U[n({\bf r})]$. This  holds -- even
mesoscopically -- for typical length scales
in excess of the inverse Fermi wavenumber $k_{\rm F}^{-1}$.
With slightly different physics, it applies exactly to
quantum-well-confined systems such as the two-dimensional
electron gas formed within heterostructure materials
(the basis of very many mesoscopic devices)
[27].

The local-density Ansatz provides a constitutive relation
to close Eq. (9a). We need another. Denoting by
${\widetilde \Delta} f^{\rm eq}$ the left-hand (mean-square)
fluctuation in the equation, we have the integral identity

$$
{\delta n({\bf r})\over \delta \mu} = {1\over k_{\rm B}T}
{\langle {\widetilde \Delta} f^{\rm eq}_{\alpha} \rangle}_{\bf k},
$$

\noindent
where we locally take the trace of
${\widetilde \Delta} f^{\rm eq}$
over the space of wavevectors ${\bf k}$.
This decouples from the spatial dependence on ${\bf r}$,
as is appropriate in a local-density treatment.
$^1$
\footnote{}{$^1$Our basic screening result does not rely on the
local-density assumption. In the wider case, Eq. (9a) reads

$$
{\widetilde \Delta} f^{\rm eq}_{\alpha}
= \Delta f^{\rm eq}_{\alpha}
{\left(
1 - \int_{\Omega} d{\bf r'}
{\delta U({\bf r})\over \delta n({\bf r'})}
{\delta n({\bf r'})\over \delta \mu}
\right)}.
$$

\noindent
Since the mean-field potential couples only to the
density, one can still integrate over $\bf k$,
leaving the real-space (albeit nonlocal) integral relation

$$
\int_{\Omega} d{\bf r'}
{\left(
\delta({\bf r} - {\bf r'})
+ {\partial n({\bf r})\over \partial \mu} 
{\delta U({\bf r})\over \delta n({\bf r'})}
\right)}
{\delta n({\bf r'})\over \delta \mu}
= {\partial n({\bf r})\over \partial \mu},
$$

\noindent
in which the extended response function
$\delta U({\bf r})/ \delta n({\bf r'})$ carries
the screening. After solving the equation
we arrive at ${\widetilde \Delta} f^{\rm eq}_{\alpha}$,
essentially as for Eq. (9b) below:

$$
{\widetilde \Delta} f^{\rm eq}_{\alpha}
= 
{\left\{
{    {\delta n({\bf r})/ \delta \mu}\over
{\partial n({\bf r})/ \partial \mu} }
\right\}}
\Delta f^{\rm eq}_{\alpha}.
$$
}
After minor rearrangement, Eq. (9a) goes to


$$
{\widetilde \Delta} f^{\rm eq}_{\alpha} =
{1\over
\displaystyle{1 +
{{\langle \Delta f^{\rm eq}_{\alpha} \rangle}_{\bf k}
\over k_{\rm B}T}{dU({\bf r})\over dn}} }
\Delta f^{\rm eq}_{\alpha}
< \Delta f^{\rm eq}_{\alpha}.
\eqno(9b)
$$

\noindent
The leading factor is the inverse static dielectric function.
(In its general form it can be seen in the final
equation of the preceding footnote.)
The total mean-square number fluctuation, in the presence
of self-consistent screening, is

$$
{\widetilde \Delta} N = \int_{\Omega} d{\bf r}
{\langle {\widetilde \Delta} f^{\rm eq}_{\alpha} \rangle}_{\bf k}.
$$

The screened analog to Eq. (7) can now be stated:

$$
\kappa \equiv {\Omega\over N^2} {\delta N\over \delta \mu}
= {\Omega\over N k_{\rm B}T} {{\widetilde \Delta N}\over N}. 
\eqno(10)
$$

\noindent
Omitting the relevant proof, already fully detailed in
Ref. [17],
we state the major outcome for nonequilibrium mesoscopic
conductors:

\item{$\bullet$}{\it In the presence of screening,
the invariant compressibility of an open, nonuniform mesoscopic
conductor -- driven at any current -- is given by Eq. (10).}

\noindent
Equation (10) stands in relation to the
random-phase approximation (RPA)
[4]
as Eq. (7) stands in relation to Lindhard. There are
differences, however.

\item{(a)} The conductor must be {\it inhomogeneous}. Otherwise
$U({\bf r}) = 0$ everywhere, and $\kappa = \kappa_{\rm free}$.
This is merely the reflection of a known fact
[4]:
in a perfectly uniform system, long-range RPA
screening is not manifested.

\item{(b)}
The conductor must be {\it degenerate}. Classically,
the Maxwell form of $f^{\rm eq}$ entails equipartition of
energy. The Coulomb term in the distribution decouples
exactly from the kinetic one,
leading to the relation ${\widetilde \Delta} N = \Delta N$.
Again, $\kappa = \kappa_{\rm free} = \kappa_{\rm cl}$.

\topinsert
\input psfig.sty
\centerline{\hskip10mm\psfig{figure=cmt25_fig2.ps,height=10.0truecm}}
\vskip -2.0truecm 

\noindent
{\bf Figure 2.}
Compressibility of a quantum-confined
two-dimensional (2D) electron gas, normalized to that
of the 2D classical gas and plotted
as a function of sheet electron density.
Dashed line: the compressibility of free conduction
electrons decreases with increasing density, owing to
their degeneracy (see Eq. (8) in the text).
Full line: the self-consistent compressibility of
interacting 2D electrons (Eq. (10) in text). With increasing
density, the mean-field Coulomb energy of a confined carrier
population will increase. That makes them collectively more
resistant to compression over and above $\kappa_{\rm free}$,
and provides a prime example of a
mesoscopic {\it many-body} effect. Since electron fluctuations
scale with $\kappa$, mesoscopic noise must directly
manifest these (and other) many-body effects.
\vskip 8truept
\endinsert
\hsize = 15.5truecm

An accessible and striking example of screening-induced
suppression occurs in the quasi-two-dimensional electron gas
(2DEG) confined within a heterojunction quantum well. In Fig. 2
we display the compressibility of a 2DEG at room temperature,
as found in any production-grade AlGaAs/InGaAs/GaAs device.
At sheet densities higher than $10^{11}$ electrons per cm$^2$,
we see the rapid onset of the Coulomb suppression in $\kappa$
where, by comparison, degeneracy acts rather more mildly.

The electrons, firmly held within the (engineered)
quantum well, exhibit their strong mutual repulsion via the
density dependence of their Hartree potential $U[n({\bf r})]$.
The consequent reduction of carrier fluctuations by
self-consistent feedback is an important element of practical
heterojunction-device design
[27].

Exactly the same Coulomb reduction of $\Delta f$
must also reveal itself in the nonequilibrium
current {\it noise}, since the same electron-hole
pair processes that underpin the compressibility
do so equally for noise
[17,20].
We explore this in the next Section.
Before doing so, let us add a novel twist to the
physics of nonequilibrium compressibility.

\eject
\vskip 8 truept
\centerline{\it d. Exchange and Correlation}
\vskip 8 truept

So far, we have treated the electron gas regardless of
exchange-correlation processes. Already in this Section,
we have seen the many-body nature of fluctuation
physics come to the fore in the compressibility response
of a nonequilibrium conductor. The introduction of
exchange-correlation corrections is not a trivial
matter; it has many levels of refinement. Here, to
give the flavor of what could be achieved for transport,
we will discuss these effects in terms of an additional
potential $U_{\rm xc}[n({\bf r})]$.

In density-functional theory, the exchange-correlation
potential augments the Hartree term. The total
compressibility can be written by inspection from
Eq. (9b):

$$
\kappa_{\rm tot} = {\Omega\over N^2 k_{\rm B}T}
\sum_{\alpha} {\Delta f^{\rm eq}_{\alpha}\over
\displaystyle{1 +
{{\langle \Delta f^{\rm eq}_{\alpha} \rangle}_{\bf k}
\over k_{\rm B}T}
{\left(
{dU\over dn}({\bf r}) + {dU_{\rm xc}\over dn}({\bf r})
\right)}
} }.
\eqno(11)
$$

\noindent
At high densities the Hartree, or RPA, term $U$
dominates. In the low-density limit the counterbalancing
exchange-correlation term $U_{\rm xc}$ is expected to dominate.
An interesting density regime should then exist,
in which the nonuniformity
of, say, a strongly confined quantum channel competes
against the shorter-ranged part of the electron-electron
interaction (with exchange). As far as we know, none of
these phenomena has been discussed systematically in
the specific context of nonequilibrium transport.

We remark that the above approach, via an
expression for $U_{\rm xc}$ which may be quite simple
(and certainly static),
has much in common with the local-field correction
long familiar in electron-gas theory
[28]. As Eq. (11) shows, our procedure
amounts to embellishing the RPA
with a term that undoes the RPA's over-correction
of the internal energy of the system; {\it quasiparticle
renormalization} never enters explicitly. Indeed,
the seminal quantum formulation of Boltzmann transport
by Kadanoff and Baym
[18]
(which informs many nonequilibrium kinetic theories,
including ours)
neglects the interaction corrections to single-particle
propagation,
while keeping particle-particle collision terms that
are of comparable order in the coupling strength.

The result of this partial inclusion of exchange-correlation
is a fully conserving and calculable
description, but one that is conceptually incomplete
(more obviously so at low carrier densities).
In a mean-field treatment, the Boltzmann--Kadanoff-Baym
collision terms will yield something close to the
exchange-correlation potential $U_{\rm xc}$.
However, the corresponding self-energy effects are not there.
$^2$
\footnote{}{$^2$
An interim fix for this incompleteness is to set the effective
mass of the carriers in the model to that of the renormalized
quasiparticles
at the Fermi surface. However, that parameter itself depends on the 
Landau quasiparticle parameter $F^s_1$ which is, by definition,
the {\it p}-wave coefficient of the two-body exchange-correlation
interaction. Again we discern the quite intricate self-consistency
between quasiparticle renormalization and collision properties.
A beautiful relation, the {\it forward-scattering sum rule}
[see e.g. W. F. Brinkman, P. M. Platzman, and T. M. Rice,
Phys. Rev. {\bf 174}, 495 (1968)],
ties together all of the Landau partial-wave amplitudes, including the
determinant of the effective mass $F^s_1$, and the {\it s}-wave term
$F^s_0$ which governs the exchange-correlation component
of the compressibility
[4].
This sum rule is not an expression of conservation, but
rather of the fermion antisymmetry that quasiparticles
inherit from the underlying ``free''-electron modes.}

It is outside our present brief to look into the
complexities of nonequilibrium-transport effects induced
by quasiparticle renormalization, and how to
include them consistently (let alone their yet-to-be-discussed
mesoscopic implications). The inclusion of renormalization
within a kinetic equation of Boltzmann type,
in a microscopically gauge-invariant way, is a challenging and
still unfinished theoretical project. It is the subject of
intensive thought by specialists. We can do no better than to
cite a small subset of this work. Several excellent
summaries, with appropriate references, appear in a volume
on many-body kinetics collated by Bonitz
[29].
Out of many detailed studies, two that address
different facets of these problems are Refs. [30] and [31].

\vskip 28 truept 
\centerline{\bf 4. NONEQUILIBRIUM FLUCTUATIONS: NOISE}
\vskip 12 truept

In the previous Section we reviewed the essentially
{\it many-body} fluctuation origin of a central result:
the compressibility sum rule for the electron gas.
We have discussed how all the quantum correlations
that condition this constraint are faithfully retained
when a system of (fluctuating) conduction electrons is

\item{(1)} {\it open} to large external reservoirs that uniquely
fix its equilibrium state, and

\item{(2)} {\it driven} out of equilibrium by external current
sources (or, equally well, by external EMFs) in a way that
always preserves global gauge invariance for the open conductor.

\noindent
The self-consistent structure of the compressibility
rule is intimately tied to charge conservation. This internal
consistency must therefore be preserved even in an approximate
transport description. {\it A model that violates the compressibility
sum rule is an incorrect model.}

Efforts to set up a sum-rule preserving transport theory
would be purely academic if the nonequilibrium fluctuations
of the electron gas could not be probed effectively in the laboratory.
While conductance is readily measured, the information that it
carries on electron-hole fluctuations is very coarse-grained. To
study the fluctuation structure, one must access the
electrical noise generated within the device. If our arguments
are correct, the noise leads directly to significant
many-body effects that are invisible in the conductance. 

One of the first well-controlled measurements of nonequilibrium
noise in a 1D mesoscopic channel was the quantum-point-contact
experiment of Reznikov {\it et al.}
[14].
Among several phenomenological treatments of this system,
the best known is the Landauer-B\"uttiker model
[2].
It predicts a noise-power spectrum $S$ that is
dominated by shot noise.

The predicted shot noise has a well developed sequence of
peaks and troughs as the channel's carrier density is swept from the
``pinch-off'' (zero-density) region, upwards through the
region where the Fermi level successively accesses the quantized
conduction subbands of the QPC.
At the same time, the conductance $G$ increases in a series
of more-or-less delineated stepwise mesas. Ideally, the steps in
$G$ come in integer units of the Landauer value $e^2/\pi\hbar$
(empirically, they hardly ever do so).

It is normal experimental practice to measure
$G$ and $S$, as functions of
carrier density, for fixed values of the source-drain
driving voltage. Although the fit to experiment is
by no means perfect, notably for noise, the standard
noise models
[2]
give a semiquantitative account of the observations.
A major innovation introduced by Reznikov {\it et al.} was
to obtain an additional series of noise data for
fixed values of source-drain {\it current}. Here,
a glaring discrepancy immediately stands out,
between the data and the standard noise prediction.

\topinsert
\vskip -0.30truecm
\input psfig.sty
\centerline{\hskip10mm\psfig{figure=cmt25_fig3.ps,height=10.0truecm}}
%
\vskip -2.20truecm 

\noindent
{\bf Figure 3.}
Low-frequency excess noise spectrum $S$ of a quantum point
contact as described in Ref. [14],
measured at fixed source-drain current $I = 300 {\rm nA}$
and plotted
as a function of carrier density in the first subband.
The density is regulated by the gate potential $V_{\rm g}$.
Full line: $S$ (with threefold enlargement).
At the subband threshold, where the Fermi level
first accesses the subband, $S$ peaks strongly.
Dot-dashed line: the outcome of Landauer-B\"uttiker noise theory
[2]
using the measured conductance.
This strictly monotonic prediction totally misses the peak.
When the channel is depleted, the carriers become classical;
there the plateau in $S$ is qualitatively similar to shot
noise, but falls short of the classical asymptotic value $2eI$.
\vskip 8truept
\endinsert
\hsize = 15.5truecm

In Fig. 3 we show a single trace of $S$ as published by
Reznikov {\it et al.}
[14],
straddling the first subband. It is evident that
the prediction of a strictly monotonic $S$ by the
Landauer-B\"uttiker noise theory
totally misses the strong peak structure in the data.
The authors of the experiment note
this. However, they make no further comment on
what is a wholly
unexpected -- and unexplained -- outcome.

There is another, purely kinetic-theoretical
explanation for the Reznikov {\it et al.} peak anomalies.
Here we sketch it in the formulation of Thakur and
the present writers
[32].
Our model, which is strictly conserving, obeys the
compressibility sum rule by construction
(for further details see Refs. [20] and [23]).

From a standard kinetic Boltzmann equation for
the carrier distribution $f(t)$, we generate an
associated equation of motion for the dynamic
mean-square fluctuation $\Delta f(t)$
[16,17]. The full 1D solution, obtained analytically
in the collision-time approximation, allows us
to write the exact nonequilibrium noise.

Normalized to the classical shot-noise expression $2eI$
at source-drain current $I$,
the free-electron excess noise (at low frequency) is 

$$
{S\over 2eI} = {\kappa\over \kappa_{\rm cl}}
{eI\over Gm^*L^2}
{\left( \tau^2_{\rm in}
+ 2{\tau_{\rm el}\tau^2_{\rm in}\over {\tau_{\rm el}+\tau_{\rm in}}}
- {\tau^2_{\rm el}\tau^2_{\rm in}\over (\tau_{\rm el}+\tau_{\rm in})^2 }
\right)}.
\eqno(12)
$$

\noindent
The set of parameters is as follows. The classical compressibility
$\kappa_{\rm cl} = 1/n k_{\rm B}T$ is in terms of the
subband carrier density $n$, keeping only the lowest subband
of the QPC as in [14].
The device conductance is $G$, while
the effective electron mass is $m^*$
and $L$ is the operational length of the 1D channel.
The inelastic and elastic collision times are, respectively,
$\tau_{\rm in}$ and $\tau_{\rm el}$.
Since the sample is {\it ballistic} -- at least in the
neighborhood of equilibrium -- the mean free path
associated with $\tau_{\rm el}$ is essentially the
ballistic device's length, $L$. Phonon emission,
however, will cause the inelastic mean free path to
become progressively shorter than $L$ as the current
intensifies.


Equation (12) has two principal features. First, the
{\it compressibility} enters directly into the nonequilibrium
excess spectrum. At high density,

$$
{\kappa\over \kappa_{\rm cl}} \to {k_{\rm B}T\over 2\varepsilon_{\rm F}}
$$

\noindent
where $\varepsilon_{\rm F}$ is the Fermi energy.
Even far from equilibrium, the excess noise necessarily
scales with the ambient temperature. The greater the degeneracy,
the smaller $S$ becomes owing to the factor above.
At low density, the electrons are classical and

$$
{\kappa\over \kappa_{\rm cl}} \to 1.
$$

\noindent
In the classical limit the factor is independent of temperature.

Second, the structure of the last right-hand term in Eq. (12)
makes $S$ very sensitive to the ratio of
the two collision times. If, as expected, $\tau_{\rm in}$
decreases substantially with the applied current $I$, the
magnitude of the excess noise will be correspondingly reduced.
This expectation provides the basis for our simulation of the result
quoted in Fig. 3. It is displayed in Fig. 4.

\topinsert
\vskip -0.30truecm
\input psfig.sty
\centerline{\hskip10mm\psfig{figure=cmt25_fig4.ps,height=10.0truecm}}
%
\vskip -2.00truecm

\noindent
{\bf Figure 4.}
Kinetic-theoretical simulation of the excess noise spectrum
of a quantum point contact, corresponding to the conditions
of Fig. 3. Density is controlled by the chemical potential
$\mu$.
Full line: $S$ (with threefold enlargement).
At the subband threshold our strictly conserving
solution for $S$ shows a strong peak that is
{\it quantitatively close} to the observed structure. 
Dot-dashed line: the outcome of Landauer-B\"uttiker noise theory
[2],
using our kinetically computed conductance.
This purely monotonic prediction still misses the peak.
Close to depletion of the channel, our simulation does not exhibit
a plateau as seen in Fig. 3. We ascribe this to the
absence of the physically quite distinct shot-noise
contribution in our present calculation, which
addresses only the excess {\it thermal} noise contribution to $S$.
\vskip 8truept
\endinsert
\hsize = 15.5truecm

Figure 4 shows our direct computation of $S$ from Eq. (12), for
environmental variables chosen to match the data of Fig. 3.
We parametrize the inelastic time as a function of $I$ and $n$
to account for the enhanced rate of phonon release. This
is then fed into Eq. (12). The elastic collision time, being
set by impurity scattering only, remains unaffected throughout.

We see a strong peak structure in the noise.
It arises from the competition among $\kappa$, $I$,
and $\tau_{\rm in}(I, n)$.
The corresponding Landauer-B\"uttiker curve
holds no trace of any peak behavior.

It is important to keep in mind that the
Landauer-B\"uttiker approach
badly violates the compressibility
sum rule, and indeed charge conservation
[20].
Our standard kinetic method respects both.
Violation of gauge invariance is built right
into the Landauer-B\"uttiker noise formula.
That is known even to its architects
[2].

\vskip 28 truept 
\centerline{\bf 5. SUMMARY}
\vskip 12 truept

In this review we have stressed that the understanding
of nonequilibrium mesoscopic transport is
indivisible from its microscopic origins in the electron gas.
The degenerate electron gas is, of course,
a prime example of a {\it multi-particle} system.

We first discussed the boundary-condition physics
of an open mesoscopic conductor. We did this with reference
to a series of searching works by a number of authors.
They are works that deserve a wider
audience -- and a far more seriously engaged one -- than
the present arbiters of mesoscopic fashion
seem able, or perhaps willing, to muster.
Certainly the papers we have recalled provide,
with considerable power, a prelude for the microscopic
analysis of open systems which we have applied
in our own nonequilibrium investigations.

Within the kinetic paradigm, we discussed how the well-known
{\it compressibility sum rule} for the electron gas
is rigidly satisfied, not just at equilibrium but far from it.
This result governs the structure of nonequilibrium
fluctuations in the driven system.
Through the action of the compressibility,
the noise spectrum of a mesoscopic wire becomes
the distinctive signature of Fermi-liquid correlations
in low-dimensional, nonequilibrium metallic-electron transport.

From the compressibility, much more might be
learned about low-dimensional current correlations.
We illustrated our fundamental many-body approach
with the graphic example of
high-field noise in a one-dimensional metallic conductor.
Its spectrum is remarkable for
its sensitivity to many-body effects, and especially for
the tight way in which these are orchestrated
by microscopic conservation.

It is obvious that any description of mesoscopic
noise, especially one that claims to be well formulated,
must always conserve charge and particle number.
There is little here to soften the implications
for other approaches to mesoscopics
[2,15,21,22].
The absolutely indispensable action of conservation
is plain. So is the direct, and by now documented,
evidence of its violation
[20].
Moreover the specific predictions of microscopic
kinetics, notably for the nonequilibrium
noise of a quantum point contact,
are a significant element in clarifying the
discussion through new experiments.

An expanded program to measure nonequilibrium noise
in quantum point contacts would be an excellent
place to look for a breakdown of nonconserving models.
It would also verify the microscopic alternative, whose
conserving properties are necessary but not sufficient for validity.
The detailed exploration of electron-electron
correlations, through renormalization of the nonequilibrium-noise
spectrum, poses an intriguing and (to our knowledge) untouched
possibility. Much has to be done to realize that potential.

\eject 
\centerline{\bf ACKNOWLEDGMENT}
\vskip 12 truept

In loving appreciation of Marjorie Ann Osborne, 1955--2002:
sister-in-law and manuscript facilitator par excellence. FG.

\vskip 28 truept
\centerline{\bf REFERENCES}
\vskip 12 truept

\item{[1]}
D. K. Ferry and S. M. Goodnick,
{\it Transport in Nanostructures}
(Cambridge University Press, Cambridge, 1997).

\item{[2]}
Ya. M. Blanter and M. B\"uttiker, {\it Phys. Rep.} {\bf 336}, 1 (2000).

\item{[3]}
N. Agra\"{\i}t, A. Levy Yeyati, and J. M. van Ruitenbeek,
{\it Preprint} cond-mat/0208239.

\item{[4]}
D. Pines and P. Nozi\`eres,
{\it The Theory of Quantum Liquids}
(Benjamin, New York, 1966).

\item{[5]}
P. Nozi\`eres, {\it Theory of Interacting Fermi Systems}
(Benjamin, New York, 1963).

\item{[6]}
A.A. Abrikosov, {\it Fundamentals of the Theory of Metals}
(North-Holland, Amsterdam, 1988).

\item{[7]}
Recall, for example, the Ward-Pitaevski-Takahashi
identities which tie the renormalization of single-particle dynamics
to the two-particle correlation vertex. The identities are the
outworking of microscopic conservation
[5];
they also entail the
{\it sum rules}
[4,5]
that must be satisfied by the electron-hole
fluctuations, and thus inevitably the current noise, of an electron gas.
See also Ref. [20] below.

\item{[8]}
F. Sols, {\it Phys. Rev. Lett.} {\bf 67}, 2874 (1991).

\item{[9]}
E. W. Fenton,
{\it Phys. Rev. B} {\bf 46}, 3754 (1992).

\item{[10]}
E. W. Fenton,
{\it Superlattices and Microstruct.} {\bf 16}, 87 (1994).

\item{[11]}
W. Magnus and W. Schoenmaker,
{\it J. Math. Phys.} {\bf 39}, 6715 (1998).

\item{[12]}
W. Magnus and W. Schoenmaker,
{\it Phys. Rev. B} {\bf 61}, 10883 (2000).

\item{[13]}
A. Kamenev and W. Kohn,
{\it Phys. Rev. B} {\bf 63}, 155304 (2001).

\item{[14]}
M. Reznikov,  M. Heiblum, H. Shtrikman, and D. Mahalu,
{\it Phys. Rev. Lett.} {\bf 75} 3340 (1995).

\item{[15]}
Y. Imry and R. Landauer, {\it Rev. Mod. Phys.} {\bf 71}, S306 (1999).

\item{[16]}
F. Green and M. P. Das,
{\it J. Phys.: Condens. Matter} {\bf 12}, 5233 (2000).

\item{[17]}
F. Green and M. P. Das,
{\it J. Phys.: Condens. Matter} {\bf 12}, 5251 (2000).

\item{[18]}
L. P. Kadanoff and G. Baym,
{\it Quantum Statistical Mechanics} (Benjamin, Reading, 1962).

\item{[19]}
L. V. Keldysh, {\it Zh. Exp. Teor. Fiz.} {\bf 47}, 1515 (1964)
({\it Sov .Phys. JETP} {\bf 20}, 1018 (1965)).

\item{[20]}
F. Green and M. P. Das, in {\it Noise and Fluctuations Control
in Electronic Devices}, edited by A. A. Balandin
(American Scientific Publishers, New York, 2002), Ch. 3.

\item{[21]}
Y. Imry, {\it Introduction to Mesoscopic Physics}
(Oxford University Press, Oxford, 1997).

\item{[22]}
S. Datta, {\it Electronic Transport in Mesoscopic Systems}
(Cambridge University Press, Cambridge, 1997).

\item{[23]}
F. Green and M. P. Das,
{\it Fluctuation and Noise Letters} {\bf 1}, C21 (2001).

\item{[24]}
From this generic structure, we see that no description
of transport -- mesoscopic or otherwise -- can avoid the
need to address truly multi-particle correlations.
Not all self-styled mesoscopic theories meet this requirement
[2,15,21,22].

\item{[25]}
N. G. van Kampen, {\it Stochastic Processes in Physics and Chemistry},
revised edition (North-Holland, Amsterdam, 2001).

\item{[26]}
R. Kubo, M. Toda, and N. Hashitsume,
{\it Statistical Physics II: Nonequilibrium Statistical Mechanics},
second edition (Springer, Berlin, 1991).

\item{[27]}
C. Weisbuch and B. Vinter,
{\it Quantum Semiconductor Structures: Fundamentals and Applications},
(Academic Press, San Diego, 1991).

\item{[28]}
G. D. Mahan, {\it Many-Particle Physics} (Plenum, New York, 1990).

\item{[29]}
M. Bonitz (editor), {\it Progress in Nonequilibrium Green's Functions}
(World Scientific, Singapore, 2000).

\item{[30]}
K. Morawetz, P. Lipavsk\'y, and V. {\accent"14 S}pi{\accent"14 c}ka,
{\it Prog. Part. Nucl. Phys.} {\bf 42}, 147 (1999). 

\item{[31]}
Yu. B. Ivanov, J. Knoll, and D. N. Voskresensky,
{\it Nucl. Phys. A} {\bf 672}, 313 (2001).

\item{[32]}
J. S. Thakur, M. P. Das, and F. Green, in process.

\end

\end{document}